# Super-resolution image display using diffractive decoders

**Authors**

Çağatay Işıl[1,2,3], Deniz Mengu[1,2,3], Yifan Zhao[1,3], Anika Tabassum[1,3], Jingxi Li[1,2,3], Yi Luo[1,2,3], Mona Jarrahi[1,3], and Aydogan Ozcan[1,2,3,*]

**Affiliations**

[1]Electrical and Computer Engineering Department, University of California, Los Angeles, CA, 90095, USA

[2]Bioengineering Department, University of California, Los Angeles, CA, 90095, USA

[3]California NanoSystems Institute (CNSI), University of California, Los Angeles, CA, 90095, USA

* ozcan@ucla.edu

**Abstract**

High-resolution synthesis/projection of images over a large field-of-view (FOV) is hindered by the restricted space-bandwidth-product (SBP) of wavefront modulators. We report a deep learning-enabled diffractive display design that is based on a jointly-trained pair of an electronic encoder and a diffractive optical decoder to synthesize/project super-resolved images using low-resolution wavefront modulators. The digital encoder, composed of a trained convolutional neural network (CNN), rapidly pre-processes the high-resolution images of interest so that their spatial information is encoded into low-resolution (LR) modulation patterns, projected via a low SBP wavefront modulator. The diffractive decoder processes this LR encoded information using thin transmissive layers that are structured using deep learning to all-optically synthesize and project super-resolved images at its output FOV. Our results indicate that this diffractive image display can achieve a super-resolution factor of ~4, demonstrating a ~16-fold increase in SBP. We also experimentally validate the success of this diffractive super-resolution display using 3D-printed diffractive decoders that operate at the THz spectrum. This diffractive image decoder can be scaled to operate at visible wavelengths and inspire the design of large FOV and high-resolution displays that are compact, low-power, and computationally efficient.

# MAIN TEXT

## Introduction

In the past decade, augmented/virtual reality (AR/VR) systems have attracted tremendous interest aiming to provide immersive and enhanced user experiences in a vast range of areas, including, e.g., human-computer interactions, visual media, art, and entertainment consumption as well as biomedical applications and instrumentation (*1*). However, the realizations of AR/VR systems have mostly relied on fixed focus stereoscopic display architectures that offer partially limited performance in terms of power efficiency, device form factor, and support of natural depth cues of the human visual system (*2–5*). Holographic displays that use spatial light modulators (SLMs) and coherent illumination with, e.g., lasers, constitute a promising alternative that allows precise control and manipulation of the optical wavefront enabling simplifications in the optical setup between the SLM and the human eye (*3*, *6*, *7*). Furthermore, this approach can emulate the wavefront emanating from a desired 3D scene to provide the depth cues of human visual perception, potentially eliminating the sources of user discomfort associated with the fixed focus stereoscopic displays e.g., vergence-accommodation conflict (*8*, *9*).

Despite these advantages, holographic displays have in general relatively modest space-bandwidth products (SBP) due to the limitations of the current wavefront modulator technology, which is directly dictated by the number of individually addressable pixels on the SLM (*7*, *10*). As a result, the current holographic display systems fail to fulfill the spatiotemporal requirements of AR/VR devices due to the limited size of the synthesized images and the extent of the corresponding viewing angles. In fact, earlier research on the subject showed that a wavefront modulator in a wearable AR/VR device must have ~50K × 50K pixels (*11*), ideally with a pixel pitch smaller than the wavelength of the visible light. Such an SLM is beyond the reach of current technology, considering that the state-of-the-art SLMs can offer resolutions up to 4K (e.g., 3,840 horizontal and 2,160 vertical pixels), with a pixel pitch that is typically 5-to-20-fold larger than the wavelength of the light in the visible part of the spectrum. Even if new SLM architectures were to be developed to meet such large SBPs, they would possibly present other challenges in terms of power consumption, memory usage, computational burden, form factor, and system complexity.

Over the years, considerable effort has been devoted to increasing the SBP of SLM technology to unleash the full potential of holographic displays, including various designs that use spatial-multiplexing of wavefront modulators arranged in application-specific configurations (*12–16*). While these multiplexed systems offer significantly larger SBPs compared to a single SLM, the utilization of multiple SLMs results in bulky optical architectures with tedious alignment and synchronization procedures in addition to increased power consumption, memory usage, and computational burden. Besides spatial-multiplexing, numerous time-multiplexing methods have also been developed for increasing the SBP of holographic displays (*17–19*), often relying on rotating mirrors and/or other moving optomechanical components, which complicate the optical setup. An alternative method of enhancing the SBP of holographic displays without any spatial- and/or time-multiplexing was presented by *Yu et al.* (*20*), where the authors introduced a complex modulation medium, e.g., multiple random diffusers, into the path of the optical signals and exploited random speckle patterns generated due to multiple light scattering events by exciting only a handful of optical modes based on wavefront shaping. While this approach provides relatively large viewing angles, the attainable image quality is deteriorated due to the random nature of the diffusers, resulting in background noise and speckle. A similar approach was also developed for AR displays by introducing periodic gratings, instead of random diffusers, into the light path between the SLM and the lens, serving as an eyepiece (*21*).

Recently, advances in machine learning have been extended to bring deep learning-enabled solutions to some of the earlier discussed challenges associated with holographic displays. Various deep neural network architectures were used to learn the transformation from a given target image to the corresponding phase-only pattern over the SLM, aiming to replace the traditional iterative hologram computation algorithms with faster and better alternatives (*22–25*). Deep neural networks have also been utilized to parameterize the wave propagation models between the SLM modulation patterns and the synthesized images for calibrating the forward model to partially account for physical error sources and aberrations present in the optical set-up (*26, 27*).

Here, we present a deep learning-enabled diffractive super-resolution (SR) image display framework (Fig. 1) based on a pair of jointly-trained electronic encoder and all-optical decoder that projects super-resolved images at the output while maintaining the size of the image field-of-view (FOV), thereby surpassing the SBP restrictions enforced by the wavefront modulator or the SLM. This diffractive SR display also enables a significant reduction in the computational burden and data transmission/storage by encoding the high-resolution images (to be projected/displayed) into compact, low-resolution representations with lower number of pixels per image, where k > 1 defines the SR factor that is targeted during training of the diffractive SR image display. In this computational image display approach (Fig. 1), the main functionality of the electronic encoder network (i.e., the front-end based on a convolutional neural network, CNN) is to compute the low-resolution (LR) SLM modulation patterns by digitally pre-processing the high-resolution images to encode LR representations of the input information. The all-optical decoder of this SR display is implemented through a passive diffractive network (*28, 29*) that is trained jointly with the electronic encoder CNN to process the input waves generated by the SLM pattern, and project a super-resolved image by decoding the encoded LR representation of the input image. Stated differently, the all-optical diffractive decoder (Fig. 1) achieves super-resolved image projection at its output FOV by processing the coherent waves generated by the LR encoded representation of the input image, which is calculated by the jointly-trained encoder CNN. This diffractive decoder forms the all-optical back-end of the SR image display, and it does not consume power except for the illumination light of the low-resolution SLM and computes the super-resolved image instantly, i.e., through the light propagation within a thin diffractive volume.

We demonstrate the SR capabilities of this unique diffractive display design using a lens-free image projection system as shown in Fig. 1. Our analyses indicate that the presented diffractive SR display can achieve an SR factor of ~4, i.e., a ~16-fold increase in SBP, using a 5-layer diffractive decoder network. We also experimentally demonstrate the success of this diffractive SR display framework based on 3D-fabricated diffractive decoders that operate at the THz part of the spectrum. This diffractive SR image display design can be scaled to work at any part of the electromagnetic spectrum, including the visible wavelengths, and will inspire image display solutions with enhanced SBP, potentially forming the building blocks of next-generation 3D display technology including, e.g., head-mounted AR/VR devices.

**Results**

The operational principles and the building blocks of the presented diffractive SR image display are depicted in Fig. 1. According to the forward model described in Fig. 1, an encoder CNN is trained to extract the spatial features of a high-resolution image (to be projected) and encode this spatial information into a lower-dimensional representation with a reduced size that is equal to the physically available number of pixels on the wavefront modulator. The input beam, which is assumed to be a uniform plane wave (see Fig. 1a), is modulated by the output pattern of the encoder network on the SLM, and subsequently, the resulting waves are all-optically processed by the diffractive decoder, aiming to recover the original image at its output FOV, effectively creating a high-resolution display through all-optical super-resolution.

Figure 2 demonstrates the super-resolved image projection performance (blind testing results) of our diffractive SR display designs trained for $k = 4$, $k = 6$, and $k = 8$ SR factors in both x and y directions. The training details of these diffractive SR displays with different configurations are described in the Methods section. Note that for each case ($k = 4, 6,$ and $8$), the input and output fields-of-view, i.e., the sizes of the wavefront modulator and the output image, are kept identical and therefore, the pixel size of the wave modulators for each SR factor is given as: $k \times 0.533\lambda$ corresponding to $2.132\lambda$, $3.198\lambda$ and $4.264\lambda$ for $k = 4, 6$ and $8$, respectively. Another important design parameter besides the SR factor ($k$) is the number of the diffractive layers, $L$, used in the all-optical decoder design. Figure 2 also provides a comparison among different decoders using $L = 1, 3,$ and $5$ diffractive layers trained for SR factors of $k = 4, 6,$ and $8$. For the results shown in Fig. 2, the wavefront modulator was assumed to provide phase-only modulation of the incoming fields; the results of a similar analysis with a complex-valued SLM at the input of each diffractive decoder are also presented in Supplementary Fig. S1. Furthermore, Supplementary Fig. S2 reports the results of an amplitude-only wavefront modulator used at the encoder.

In Fig. 2 and Supplementary Figs. S1-S2, we can see that the cases with $k \geq 4$ describe a very low-resolution SLM with a large pixel size and a small number of pixels, for which the native resolution is insufficient to directly represent most of the details of the test objects (EMNIST handwritten letters) within the FOV. On the other hand, these spatial features can be recovered all-optically through the diffractive decoder, projecting SR images at its output FOV, as illustrated in Fig. 2 and Supplementary Figs. S1-S2. We also observe that, for a fixed SR factor, $k$, the discrepancies between the desired high-resolution images and the optically synthesized intensity distributions at the output FOV of the diffractive decoder become smaller as the number of diffractive layers, $L$, increases, demonstrating the advantage of deeper diffractive decoders to provide better image projection.

Beyond the visual inspections and comparisons provided in Fig. 2, Supplementary Figs. S1 and S2, the efficacy of the diffractive SR display framework is also confirmed by quantifying the image quality using the structural similarity index measure (SSIM) and the peak signal-to-noise ratio (PSNR) metrics. As part of this quantitative analysis, Figure 3 compares the overall image synthesis performance of phase-only and complex-valued wavefront modulation at the input plane of the diffractive decoders. On average, complex-valued wavefront modulation provides slightly better PSNR and SSIM values at the output of the diffractive decoder compared to the phase-only modulation/encoding because of the increased degrees of freedom. Figure 3 also supports the conclusion of Fig. 2 that the deeper diffractive decoders with a larger number of diffractive layers overall perform higher fidelity output image projection.

To provide more insights into the success of our diffractive decoders in synthesizing super-resolved images, we conducted additional blinded tests using the images of various lines with subpixel linewidths compared to the native phase-only SLM resolution, as shown in Fig. 4. It is important to emphasize that the training of our diffractive SR systems entirely relied on the EMNIST handwritten letters dataset; hence, these new images of resolution test lines represent a blind testing dataset that is statistically different from the training data. These resolution test results summarized in Fig. 4 for phase-only encoding reveal that even for deeply subpixel linewidths, the individual lines in both the horizontal and vertical structures can be resolved at the output of the 5-layer diffractive decoder; see for example, $2.132\lambda$ lines, encoded through a phase-only SLM with a pixel size of $4.264\lambda$. On the other hand, the all-optical decoder with a single diffractive layer (L=1) fails to resolve the individual lines with a linewidth of $2.132\lambda$ for k = 8 (Fig. 4) due to the limited generalization capability offered by the 1-layer diffractive decoder architecture. Supplementary Fig. S3 also illustrates the same resolution test analysis except for a diffractive SR display system using complex-valued encoding at the SLM, arriving at similar conclusions.

These results, summarized in Fig. 4, demonstrate that $2.132\lambda$ linewidth test images composed of vertical and horizontal line pairs can be resolved through the L = 5 diffractive decoder trained with an SR factor of k = 8 using a phase-only wavefront modulator with a native pixel size of $4.264\lambda$, i.e., k x $0.533\lambda$. This indicates that the effective pixel size at the output plane of this diffractive decoder is ~$1.066\lambda$ (half of the minimum resolvable linewidth (*30*)) which corresponds to a pixel super-resolution factor of ~4-fold and an SBP increase of ~16-fold. For comparison, the same resolution test target images with a linewidth of $2.132\lambda$ cannot be resolved, as expected, by low-resolution displays that have a pixel size of $2.132\lambda$ or larger, as shown in Fig. 4 right column. However, the diffractive SR display with L = 5 and k=8 successfully resolved these lines using a pixel size of $4.264\lambda$ at the phase-only wavefront encoder, corresponding to ~16-fold increase in the SBP of the image display system. We should note that this increase in the SBP is smaller than $k^2$, which indicates that the training image set (handwritten EMINST letters) did not have sufficient representation of higher resolution features to guide the joint-training of the encoder-decoder pair to achieve even higher resolution image display; furthermore, such resolution test targets composed of lines or gratings were not included in our training data.

Next, to experimentally demonstrate the success of the presented SR image display framework, we trained two different diffractive decoders designed for operation at the THz part of the spectrum (see the Methods section for details). The first one uses a 3-layer diffractive decoder design (Figs. 5-6), and the second decoder (Fig. 7) relies only on a single diffractive surface, L=1, to achieve image SR. These diffractive decoders were 3D-printed and physically assembled/aligned to operate under continuous-wave THz illumination at λ = ~0.75 mm (see the Methods section). The experimental setup, the 3D-printed diffractive decoders, and the phase profiles of the fabricated optimized diffractive layers are illustrated in Figs. 5 and 7, for the 3-layer and 1-layer diffractive decoders, respectively. As detailed in the Methods section, the training loss function of these fabricated diffractive decoders included an additional penalty term regularizing the output diffraction efficiency, which is on average 2.39% and 3.29% for the 3-layer and 1-layer decoders, respectively, for the blind test images. Furthermore, these diffractive decoders were trained to be resilient against layer-to-layer misalignments in x, y, and z directions using a vaccination strategy (outlined in the Methods section) that randomly introduces 3D misalignments during the training process, which was shown to create misalignment tolerant diffractive designs (*31*).

The experimental results of the diffractive SR image display system with L = 3 layers are shown in Fig. 6, clearly demonstrating the super-resolution capability of the diffractive decoder at its output FOV, also providing a very good match between the numerical forward model results and the experimental measurements. Similarly, Fig. 8 reports the success of our experimental results obtained using the SR image display system with a single diffractive layer (L=1), also achieving super-resolution at the output of the diffractive decoder. Despite using a single diffractive layer in the decoder, the jointly-trained encoding-decoding framework optically synthesized the target test letters at the output FOV. In these experiments, the average PSNR values achieved by the diffractive decoders are 13.134 ± 1.368 dB for L = 3 and 12.151 ± 2.138 dB for L = 1. These results are in line with our former analysis reported in Figs. 2-4, confirming the advantages of deeper diffractive decoders for better image synthesis at the output FOV.

Finally, the resilience of the SR image display framework to different quantization levels of the wavefront modulation is illustrated in Fig. 9. For this analysis, the diffractive SR image display system with L = 5 layers trained for 16-bit quantization of phase-only wavefront modulator was blindly tested for lower quantization levels at 8-, 6-, 4-, and 2-bit. Figure 9 shows that the presented diffractive SR image display can successfully synthesize super-resolved images at its output even for 6-bit quantization of the encoded phase profiles. The overall image synthesis performance of the 8-bit (18.58 dB PSNR and 0.58 SSIM) and 6-bit (18.20 dB PSNR and 0.55 SSIM) quantization of the phase modulator/encoder demonstrates the robustness of the diffractive system, considering that we have 18.61 dB PSNR and 0.58 SSIM for the 16-bit phase quantization case. The diffractive SR image display fails to synthesize clear images at its output FOV for 2-bit phase quantization, and is partially successful for 4-bit phase quantization (Fig. 9). For these lower bit-depth phase quantization cases, the presented encoding-decoding framework can be trained from scratch to further improve the image projection performance under limited phase encoding precision.

**Discussion**

We presented a diffractive SR image display framework based on a jointly-trained pair of encoder and decoder networks that collectively improve the SBP of the image projection system. The deep learning-designed diffractive display system synthesizes and projects super-resolved images at its output FOV by encoding each high-resolution image of interest into low-resolution representations with lower number of pixels per image. As a result of this, the all-optical decoding capability of the diffractive network not only improves the effective SBP of the image projection system but also reduces the data transmission and storage needs since low-resolution wavefront modulators are used. The decoder network is an all-optical diffractive system composed of passive structured surfaces and therefore does not consume computing power except for the illumination light. Similarly, the all-optically synthesized images are computed at the speed of light propagation between the encoder SLM plane and the diffractive decoder output FOV, and therefore the only computational bottleneck for speed and power consumption is at the inference of the front-end CNN encoder.

As shown in our experimental results (Figs. 6 and 8), there are some relatively small discrepancies between the numerical output images of our forward model and the corresponding experimentally measured output images. There are potential error sources that might cause these discrepancies. First, our numerical forward model used in the training assumes a uniform plane wave incident on the surface of the wavefront modulator, and this assumption could potentially be violated in our experimental setup due to wavefront distortions of the THz source used. Additional errors might have occurred during the fabrication of each diffractive layer due to the limited resolution of our 3D printer. Furthermore, any inaccuracy in the characterization of the refractive index of the 3D printing material at the illumination wavelength is yet another factor that might also be partly responsible for the small mismatch between our numerical and experimental results.

Although the THz part of the electromagnetic spectrum was used for these proof-of-concept experimental demonstrations, the main design principles and conclusions provided in our study also apply to display systems operating at visible wavelengths. Extending the presented SR display designs to visible wavelengths is feasible using various nano-fabrication techniques providing subwavelength features, e.g., two-photon polymerization and lithography (*32*, *33*). Furthermore, in this study, we investigated the capabilities of the jointly-trained encoder and decoder networks in synthesizing SR images at a small axial distance (~150-350$\lambda$) from the wavefront modulation plane of the encoder. Our training procedures and design principles can also be extended for synthesizing 3D super-resolved object fields covering an extended working distance at the output of the decoder network.

While the SR image display results of this manuscript were obtained at a single illumination wavelength, we can also extend the design principles of diffractive decoders to operate at multiple wavelengths to bring spectral information into the projected images. To optically synthesize full-color (RGB) images, some of the traditional holographic display systems use sequential operation (i.e., one illumination wavelength at a given time followed by another wavelength), which spatially utilizes all the pixels of the SLM for each wavelength at the expense of reducing the frame rate (*34*). Spatial multiplexing of the SLM pixels among different illumination wavelength channels constitutes an alternative option, although this approach further sacrifices the SBP of the display among different color channels, restricting the output image size and the resolution. By incorporating the dispersion characteristics and the refractive index information of the wavefront modulation medium (e.g., liquid-crystal) and the diffractive decoder material as part of the optical forward model of our design, the presented diffractive display design can be extended to synthesize super-resolved images at a group of illumination wavelengths. In this case, the jointly-trained encoder network can be optimized to drive the SLM at multiple wavelengths, either simultaneously or sequentially, based on the assumption made during the training process of the encoder-decoder pair. In either mode of operation, multi-wavelength SR image displays using diffractive decoders need more diffractive features/neurons for a given output FOV and SR factor compared to their monochrome versions to be able to handle independent spatial features at different illumination wavelengths or color channels of the input image.

The presented SR image display can be thought of as a hybrid autoencoder framework containing a digital encoder network that is used to create low-dimensional representations of the target high-resolution images and an all-optical diffractive decoder (jointly-trained with the encoder) to synthesize super-resolved images at its output FOV from the diffraction patterns of these low resolution encoded patterns generated by the encoder network. This joint optimization and the communication between the electronic front-end and the diffractive optical back-end of our SR image display is crucial to increase the SBP of the image formation models and will inspire the design of the new high-resolution camera and display systems that are compact, low-power, and computationally-efficient.

## Methods

### All-optical decoder design for SR image displays

In our optical forward model, the diffractive modulation layers are discretized over a regular 2D grid with a period of $w_x$ and $w_y$ for the x- and y- axes, respectively. Each point in the grid, termed 'diffractive neuron', denotes the transmittance coefficient $t_l[m, n]$ of the smallest feature in each modulation layer. The field transmittance of a diffractive layer, $l \geq 1$, is defined as:

$$t_l[m, n] = \exp\left(j\frac{2\pi}{\lambda}(\tau(\lambda) - n_a)h_l[m, n]\right)$$

(1)

where $\tau(\lambda) = n(\lambda) + j\kappa(\lambda)$ is the complex refractive index of the optical material used to fabricate the diffractive layers, $\lambda$ denotes the wavelength of the coherent illumination. $n_a = 1$ refers to the refractive index of the medium (air in our case) surrounding the modulation layers, and $h_l[m, n]$ represents the material thickness of the corresponding neuron, which is defined as

$$h_l[m, n] = \frac{\tanh(o_l[m, n]) + 1}{2}(h_m - h_b) + h_b$$

(2)

where $o_l[m, n]$ is an auxiliary input variable used to compute the material thickness values between $[h_b, h_m]$. These auxiliary variables $o_l[m, n]$ and the material thickness values $h_l[m, n]$ for all $m, n$ & $l$ are optimized using stochastic gradient descent based error backpropagation and deep learning (*35*, *36*).

The 2D modulation function $T_l(x, y)$ for continuous coordinates $(x, y)$ can be written in terms of transmittance coefficients $t_l[m, n]$ and 2D rectangular sampling kernels $p_l(x, y)$ as follows:

$$T_l(x, y) = \sum_m \sum_n t_l[m, n]\, p_l(x - m w_x, y - n w_y)$$

(3)

where $p_l(x, y)$ is defined as

$$p_l(x, y) = \begin{cases} 1, & |x| < \frac{w_x}{2}\ \&\ |y| < \frac{w_y}{2} \\ 0, & otherwise \end{cases}$$

(4)

The light propagation between successive diffractive layers is modeled by a fast Fourier transform (FFT)-based implementation of the Rayleigh-Sommerfeld diffraction integral, using the angular spectrum method. This diffraction integral can be expressed as a 2D convolution of the propagation kernel $w(x, y, z)$ and the input wavefield $U'_l(x, y)$:

$$U_{l+1}(x, y) = U'_l(x, y) * w(x, y, z_{l+1} - z_l),$$

$$U'_l(x, y) = U_l(x, y) T_l(x, y),$$

$$w(x, y, z) = \frac{z}{r^2}\left(\frac{1}{2\pi r} + \frac{1}{j\lambda}\right) \exp\left(j \frac{2\pi r}{\lambda}\right),$$

$$r = \sqrt{x^2 + y^2 + z^2}.$$

(5)

**Diffractive decoder vaccination**

To mitigate the impact of potential misalignments during the experiments, we incorporated possible physical error sources as part of the optical forward model of the diffractive decoders. During the training of the experimentally-tested diffractive designs, these errors were modeled using random 3D displacement vectors, $D^l = (D_x, D_y, D_z)$ denoting the deviation of the location of diffractive layer $l$, from its ideal position, where $D_x, D_y$, and $D_z$ were defined as uniformly distributed independent random variables,

$$D_x \sim U(-\Delta_x, \Delta_x),$$

$$D_y \sim U(-\Delta_y, \Delta_y),$$

$$D_z \sim U(-\Delta_z, \Delta_z).$$

(6)

The variables $\Delta_x$, $\Delta_y$, and $\Delta_z$ in Eq. 6 denote the maximum amount of displacement along the corresponding axis. Accordingly, the position of the diffractive layer $l$ at $i^{th}$ iteration $L^{(l,i)}$ was defined as

$$L^{(l,i)} = \left(L_x^l, L_y^l,, L_z^l,\right) + (D_x^{(l,i)}, D_y^{(l,i)}, D_z^{(l,i)}).$$

(7)

**Encoder CNN network design for SR image displays**

We used a CNN-based electronic encoder to compress high-resolution input images of interest into lower-dimensional latent representations that can be presented using a low SBP wavefront modulator or SLM. The CNN network architecture is illustrated in Fig. 1a. It contains 4 convolutional blocks, followed by a flattening operation, a fully connected layer, a rectified linear unit (ReLU) based activation function, and an unflattening operation. Each convolutional block contains 3 pairs of 4x4 convolutional filters ("same" padding configuration) with a Leaky ReLU (with a slope of α = 0.1). For the $i^{th}$ convolutional block, there are $2^{1+i}$ channels. To decrease the dimensions of the channels and obtain low-dimensional representations at the output of the electronic encoder CNN, a fully connected layer is utilized at the end.

**Training and test dataset preparation**

We created a training image dataset, namely the EMNIST *display* dataset, to train and test our diffractive SR display system. As seen in Supplementary Fig. S4, each image in this display dataset was generated by using different numbers of images selected from the EMNIST handwritten letters. The selected letters were augmented by predefined geometrical operations including scaling ($K \sim U(0.84, 1)$), rotation ($\Theta \sim U(-5,° 5°)$), and translation ($D_x, D_y \sim U(-1.06\lambda, 1.06\lambda)$). Then, these selected and augmented images were randomly placed in a 3x3 grid. This procedure was used for each image in our display dataset. In the original EMNIST handwritten letters dataset, there are 88,000 and 14,800 letter images for training and testing, respectively (*37*). The original size of these letters is 28 x 28 pixels. Before applying the tiling procedure described above, each image was interpolated to 32 x 32 using bicubic interpolation. For the training dataset, 60,000 images (96 x 96 pixels) containing 1, 2, 3, and 4 different handwritten letters (15,000 images for each case) were created using the EMNIST letters training dataset. For the validation dataset, 6,000 images (96 x 96 pixels) containing 1, 2, 3, and 4 different handwritten letters (1,500 images for each case) were created using the EMNIST letters training dataset. For the test dataset, 6,000 images (96 x 96 pixels) containing 6, 7, 8, and 9 handwritten letters (1,500 images for each case) were created using the EMNIST letters test dataset.

In the experimentally tested designs, two complementary image sets containing 80,000 and 8,800 different handwritten letters from the EMNIST letters training dataset were used as the training and validation datasets, respectively. The EMNIST letters test dataset with 14,800 different handwritten letters was used as our test dataset. The handwritten letters were resized to 15 x 15 pixels using bicubic downsampling (with an anti-aliasing filter), based on the effective pixel size used at the measurement (output) plane.

**Implementation details of the numerically-tested diffractive SR display systems**

Diffractive neuron width of the transmissive layers ($w_x, w_y$) and the sampling period of the light propagation model were chosen as $0.533\lambda$. Each diffractive layer had a size of $106.66\lambda \: x \: 106.66\lambda$ (200 x 200 pixels). The input and output FOVs of the diffractive decoders were $51.168\lambda \: x \: 51.168\lambda$ (96 x 96 pixels). To avoid aliasing in our optical forward model, these matrices were padded with zeros to have 400x400 pixels. In our optical forward propagation model, the material absorption was assumed to be zero ($\kappa(\lambda) = 0$). Therefore, the transmittance coefficient of each feature of a diffractive layer can be written as:

$$t_l[m,n] = \exp(j\theta_l[m,n]) = \exp\left(j\frac{2\pi}{\lambda}(n(\lambda) - n_a)h_l[m,n]\right).$$

(8)

Phase coefficients $\theta_l[m,n]$ of each diffractive layer of the decoder was optimized using deep learning and error backpropagation. The phase coefficients $\theta_l[m,n]$ were initialized as 0.

Different diffractive decoders, including 1, 3, and 5 transmissive layers were analyzed in our results. The axial distances (Fig. 1b) from the input plane to the first diffractive layer $d_1$, from one diffractive layer to another diffractive layer $d_2$, and from the last layer to the output plane $d_3$ were optimized empirically (see Table 1). The trained phase profiles of the transmissive layers of the diffractive decoders using phase-only SLMs are reported in Supplementary Fig. S5.

**Table 1. Axial distances for different diffractive decoder designs including 1, 3, and 5 diffractive layers used in our numerical and experimental results**

|  |  | $d_1$ | $d_2$ | $d_3$ |
|---|---|---|---|---|
| **Numerical results** | 1 layer | $6.667\lambda$ | - | $173.333\lambda$ |
|  | 3 layers | $4\lambda$ | $53.334\lambda$ | $53.334\lambda$ |
|  | 5 layers | $2.667\lambda$ | $66.667\lambda$ | $80\lambda$ |
| **Experimental results** | 1 layer | $26.667\lambda$ | - | $173.333\lambda$ |
|  | 3 layers | $26.667\lambda$ | $53.334\lambda$ | $53.334\lambda$ |

**Implementation details of the experimentally-tested diffractive SR display systems**

In our experiments, a monochromatic THz illumination source ($\lambda = \sim 0.75 \: mm$) was used. The diffractive neuron size of the transmissive layers and the sampling period of the light propagation model were chosen as $\sim 0.667\lambda$ and the size of each layer was determined to be $66.7\lambda \: x \: 66.7\lambda$ (5 cm $\times$ 5 cm). The effective pixel size at the measurement plane was selected as $\sim 2.67\lambda$ in our experiments. The size of the phase-only wavefront modulator was selected as $40\lambda \times 40\lambda$ (3 cm $\times$ 3 cm), which is also equal to the size of the output FOV. Based on the $\sim 2.67\lambda$ pixel size at the output FOV, the number of the output image pixels was set to be $15 \times 15$ and the LR wavefront modulator was selected as a $5 \times 5$ pixel phase-only layer with a pixel pitch of $8\lambda \times 8\lambda$. This corresponds to a desired SR factor of k=15/5=3 that was targeted during the training of these models i.e., a 9-fold SBP enhancement through the diffractive decoder. To avoid spatial aliasing in the optical forward model, the matrices were padded with zeros to have $300 \times 300$ pixels.

The complex refractive index of the 3D-printing material $\tau(\lambda)$ used to fabricate the diffractive layers and the phase-encoded inputs was measured as $\sim 1.6518 + j0.0612$. In our model, the material thickness of each diffractive neuron $h_l[m,n]$ was optimized in the range of $[0.5\ mm, \sim1.64\ mm]$ that corresponds to $[-\pi, \pi)$ for phase modulation. The phase coefficients $\theta_l[m,n]$ were initialized as 0.

Two diffractive decoders with L=3 and L=1 were fabricated and tested in the experiments. The axial distances for these models are given in Table 1. As part of our diffractive decoder vaccination process, independent random alignment errors along the axial and lateral coordinates were added to the positions of the diffractive layers and the input FOV during the training phase, as detailed in Eq. (7). During the training of these experimentally tested models, $\Delta_x$, $\Delta_y$, and $\Delta_z$ were set to be $\sim 0.334\lambda$, $\sim 0.334\lambda$, and $\sim 0.533\lambda$, respectively. The optimized thickness maps for the resulting diffractive layers and the phase-only encoded representations were converted into STL files using MATLAB and they were fabricated by using a 3D printer (Objet30 Pro, Stratasys Ltd.).

**Experimental setup**

The schematic diagram of the experimental setup is shown in Figs. 5c-d. The THz plane wave incident on the object was generated through a WR2.2 modular amplifier/multiplier chain (AMC) with a compatible diagonal horn antenna (Virginia Diode Inc.). The AMC received a 10 dBm RF input signal at 11.111 GHz ($f_{RF1}$) and multiplied it 36 times to generate a continuous-wave (CW) radiation at 0.4 THz. The AMC output was modulated at a 1 kHz rate to resolve low-noise output data through lock-in detection. The exit aperture of the horn antenna was placed ~60 cm away from the object plane of the 3D-printed diffractive decoder. The diffracted THz radiation at the output plane was detected with a single-pixel Mixer/AMC (Virginia Diode Inc.). A 10 dBm RF signal at 11.083 GHz ($f_{RF2}$) was fed to the detector as a local oscillator for mixing, to down-convert the detected signal to 1 GHz. The detector was placed on an X-Y positioning stage, including two linear motorized stages (Thorlabs NRT100). The output FOV was scanned using a $0.5 \times 0.25$ mm detector with a step size of 1 mm. A 2×2-pixel binning was used to increase the SNR and approximately match the output pixel size of our design, i.e., ~2.67λ. The down-converted signal was sent to cascaded low-noise amplifiers (Mini-Circuits ZRL-1150-LN+) to obtain a 40 dB amplification. Then, a 1 GHz (+/-10 MHz) bandpass filter (KL Electronics 3C40-1000/T10-O/O) was used to eliminate the noise coming from unwanted frequency bands. The amplified and filtered signal passed through a tunable attenuator (HP 8495B) for linear calibration and a low-noise power detector (Mini-Circuits ZX47-60). The output voltage signal was read by a lock-in amplifier (Stanford Research SR830). The modulation signal was used as the reference signal for the lock-in amplifier. According to the calibration results, the lock-in amplifier readings were converted to a linear scale. The bottom 5% and the top 5% of all the pixel values of each measurement were saturated and the remaining pixel values were mapped to a dynamic range between 0 and 1.

**Training loss function and performance comparison metrics**

For the joint training of an electronic CNN-based encoder and an optical diffractive decoder, the mean absolute error function together with an efficiency penalty term ($e^{-\eta}$) was used as our training loss function ($\mathcal{L}$), which is defined as:

$$\mathcal{L} = \frac{1}{N}\sum_{i=1}^{N}|y_i - \sigma\hat{y}_i| + \gamma e^{-\eta},$$

$$\sigma = \frac{\sum_{i=1}^{N} y_i}{\sum_{i=1}^{N} \hat{y}_i},$$

$$\eta = 100 \times \frac{P_o}{P_i}.$$

(9)

where $y_i$ and $\hat{y}_i$ denote the target (ground truth) high-resolution image and the all-optical decoder output intensity, respectively. $N$ represents the number of pixels in each image, and $\sigma$ is a normalization term. $P_i$ and $P_o = \sum_{i=1}^{N}\hat{y}_i$ represent the optical power incident on the input FOV and the output FOV, respectively. The power efficiency of a diffractive decoder can be adjusted by tuning $\gamma$. For the training of the experimentally demonstrated diffractive decoders, $\gamma$ was set to be 0.005 for L=1 and 0.015 for L=3; for the other designs, $\gamma = 0$.

During the joint training of the CNN encoder and the diffractive decoder, several image data augmentations, including random image rotations (0, 90, 180, and 270 degrees), random flipping of images, and random contrast adjustments were used. The joint training was implemented in Python (v3.6.12) and TensorFlow (v1.15.4, Google LLC). Adam optimizer was used during the training with a learning rate of 0.001 for the all-optical diffractive decoders and 0.0005 for the CNN-based encoders (*38*). All the networks were trained using a GeForce RTX 3090 GPU (NVidia Corp.) and an AMD Ryzen Threadripper 3960X CPU (Advanced Micro Devices Inc.) with 264 GB of RAM. We trained each network for at most 17 hours (500 epochs) with a batch size of 40.

For quantitative comparison of our results, PSNR and SSIM values were calculated for each target image in our test sets (*39*). PSNR is computed as follows

$$PSNR = 10\log_{10}\left(\frac{1}{\frac{1}{N}\sum_{i=1}^{N}|y_i - \sigma\hat{y}_i|^2}\right).$$

(10)

SSIM is computed using the standard implementation in TensorFlow with a maximum value of 1. For visual comparison, the low resolution versions of the target images were obtained using k-fold downsampling with a bicubic kernel (and an anti-aliasing filter) and were illustrated in the figures.

# Figures and Figure Captions

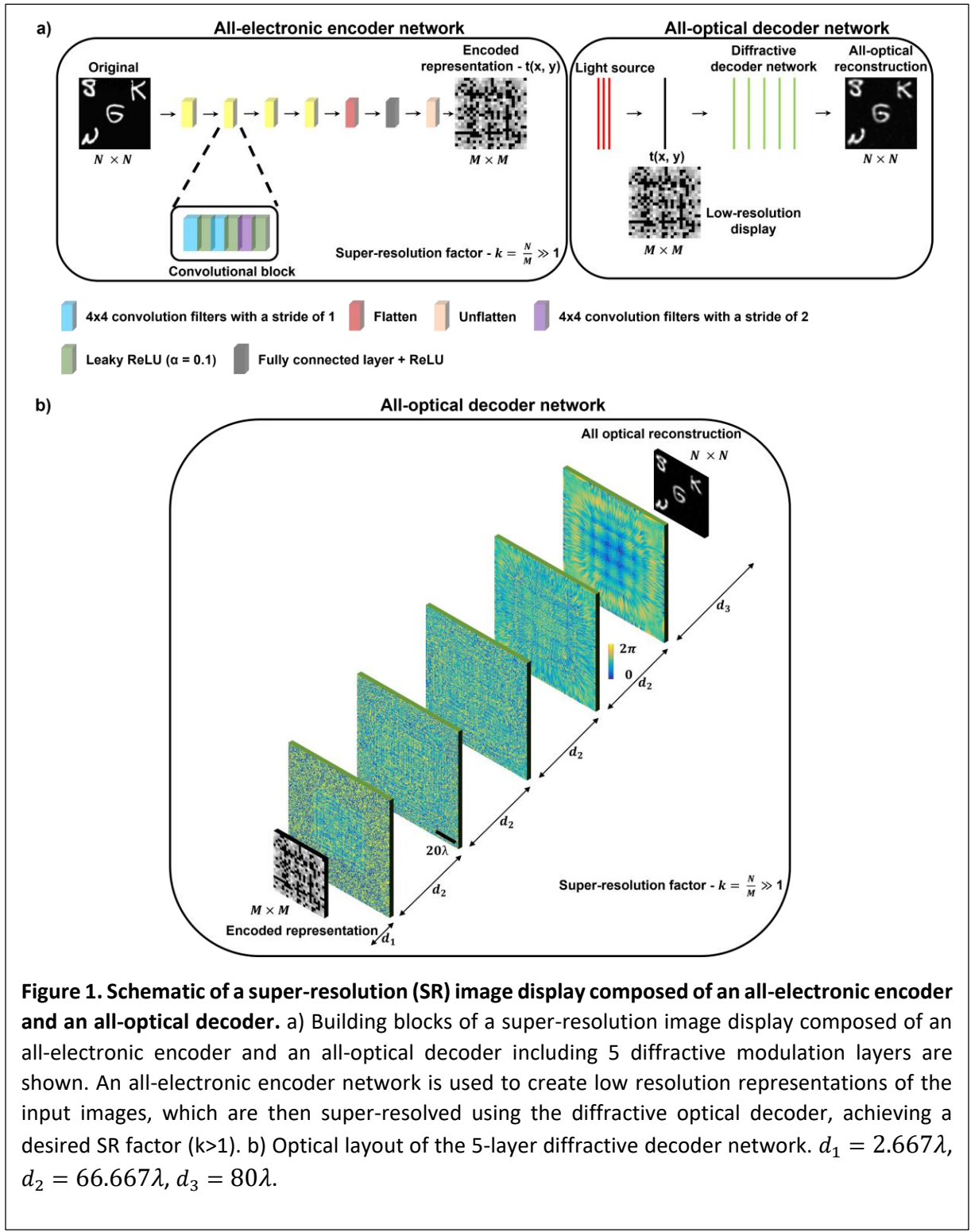

**Figure 1. Schematic of a super-resolution (SR) image display composed of an all-electronic encoder and an all-optical decoder.** a) Building blocks of a super-resolution image display composed of an all-electronic encoder and an all-optical decoder including 5 diffractive modulation layers are shown. An all-electronic encoder network is used to create low resolution representations of the input images, which are then super-resolved using the diffractive optical decoder, achieving a desired SR factor (k>1). b) Optical layout of the 5-layer diffractive decoder network. $d_1 = 2.667\lambda$, $d_2 = 66.667\lambda$, $d_3 = 80\lambda$.

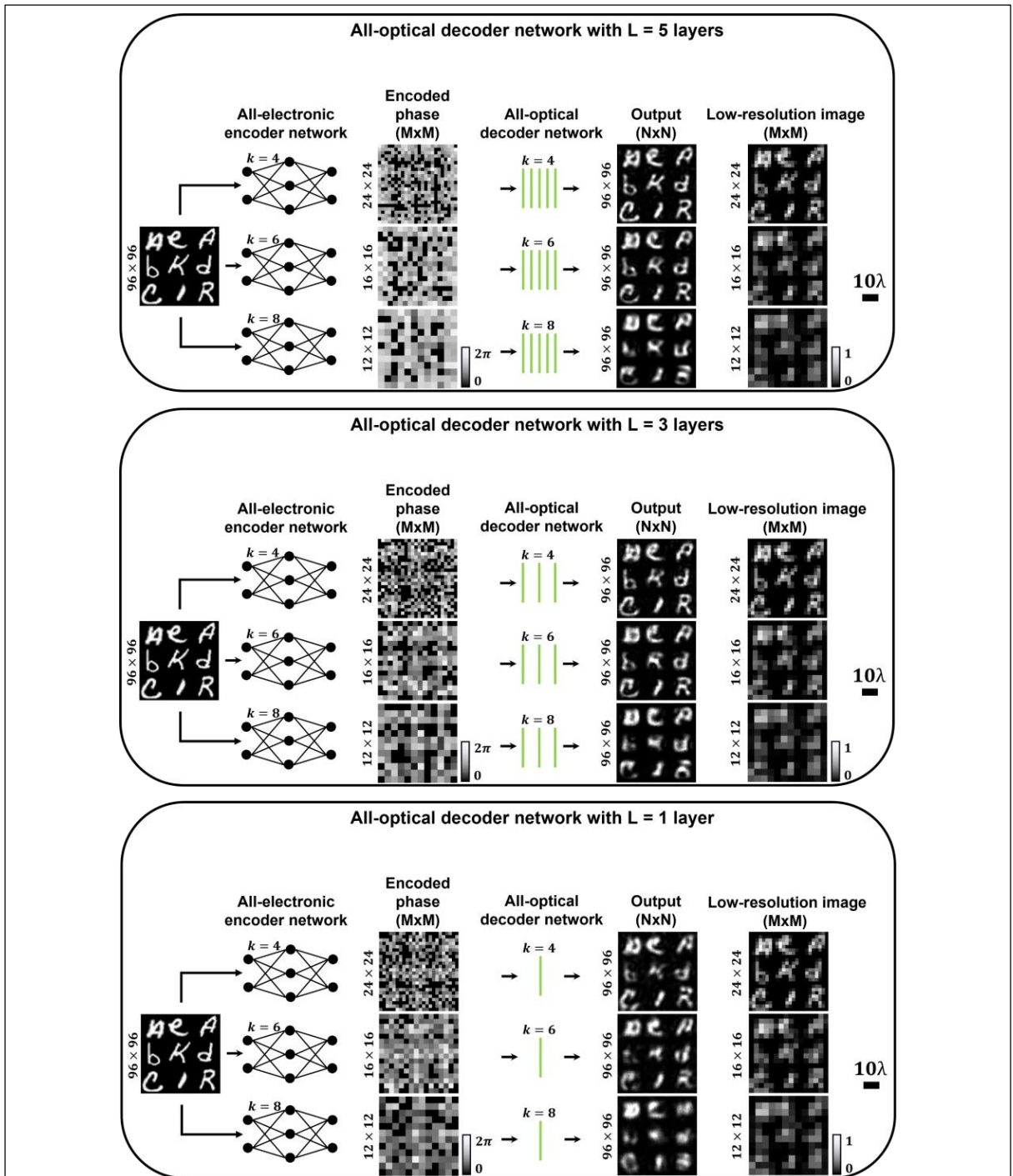

**Figure 2. Image projection results of the diffractive SR display using a phase-only SLM.** Top: Image projection results of the SR display using 5 diffractive layers (L=5). Middle row: Image projection results of the SR display using 3 diffractive layers (L=3). Bottom: Image projection results of the SR display using 1 diffractive layer (L=1). For comparison, low resolution versions of the same images using the same number of pixels as the corresponding wavefront modulator are illustrated on the right side of the figure.

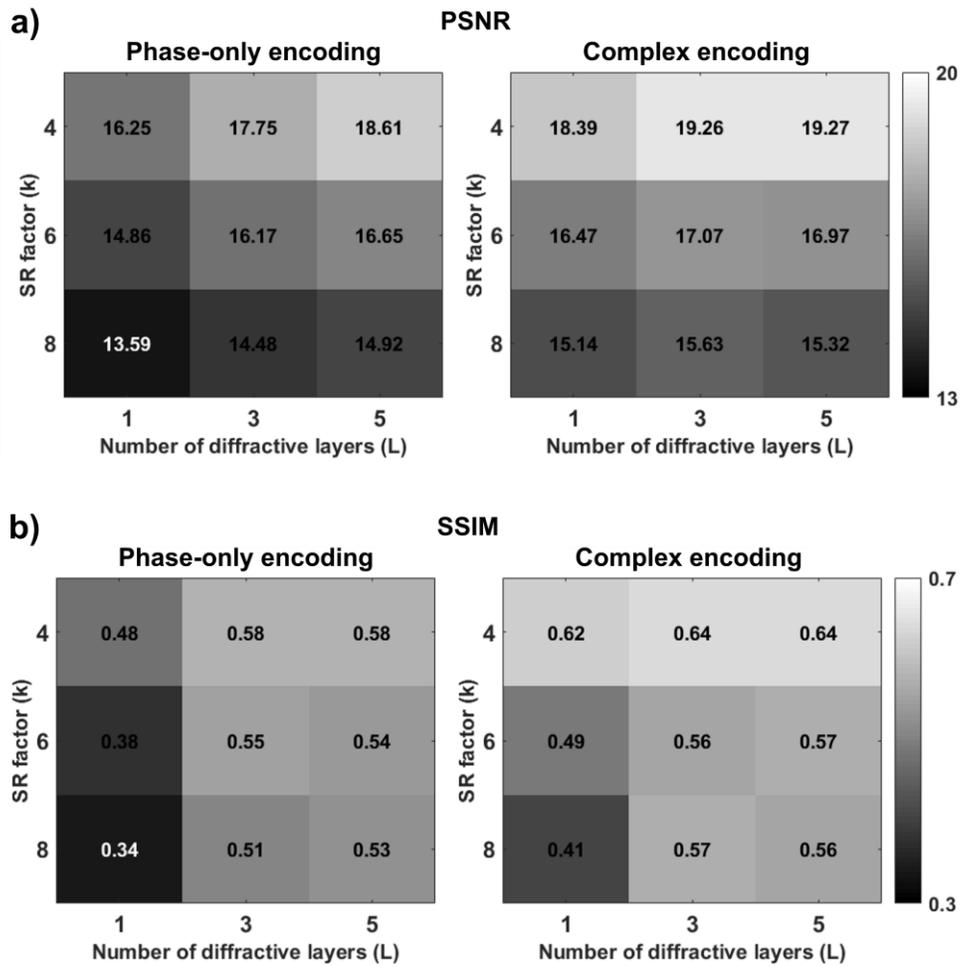

**Figure 3. Quantification of the image projection performance of the diffractive SR display as a function of k and L.** The test image dataset contains 6000 images, each containing multiple EMNIST handwritten letters. a) Average PSNR values for phase-only (left) and complex-valued (right) encoding. b) Average SSIM values for phase-only (left) and complex-valued (right) encoding.

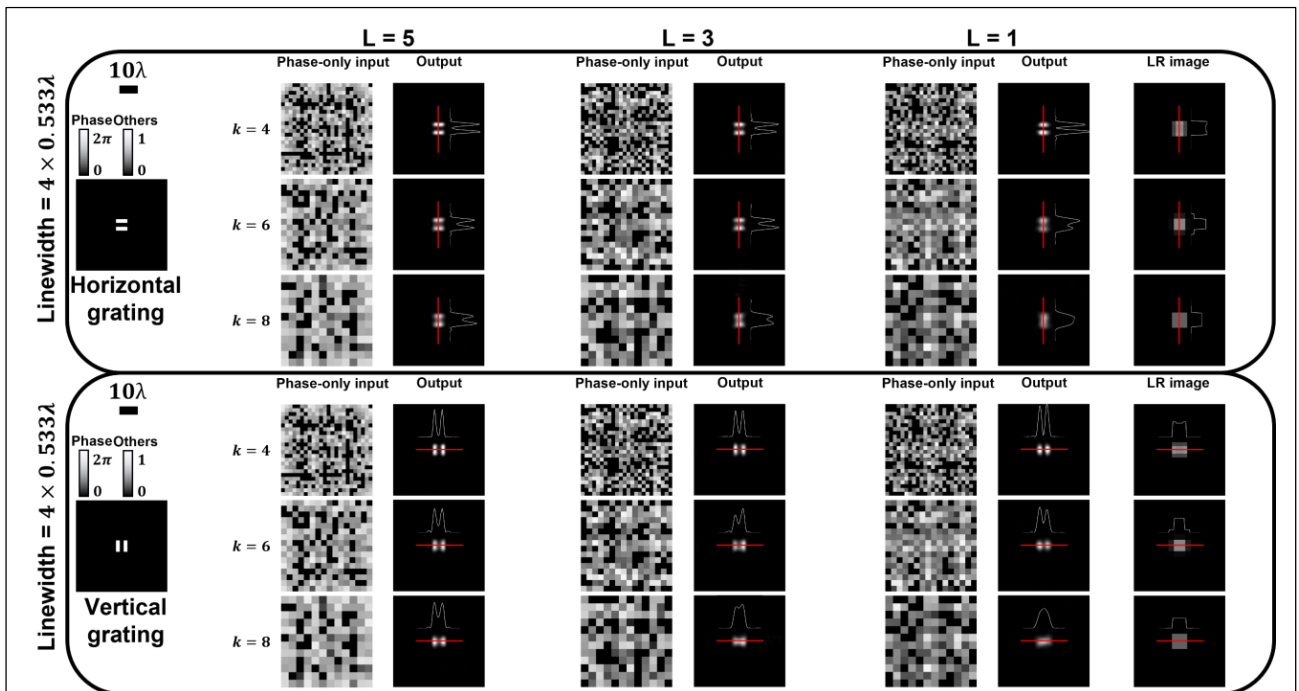

**Figure 4. Image resolution analysis of the diffractive SR display using a phase-only SLM.** Projections of vertical and horizontal line-pairs with a linewidth of $2.132\lambda$ are demonstrated for different SR factors (k = 4, 6, and 8). Diffractive all-optical decoders with different numbers of diffractive layers (L = 1, 3, and 5) project super-resolved images at the output. For comparison, low resolution (LR) versions of the same objects using the same number of pixels as the corresponding wavefront modulator are illustrated on the right side of the figure. The diffractive SR systems were trained using handwritten letters and the training dataset did not include any resolution test targets or line pairs.

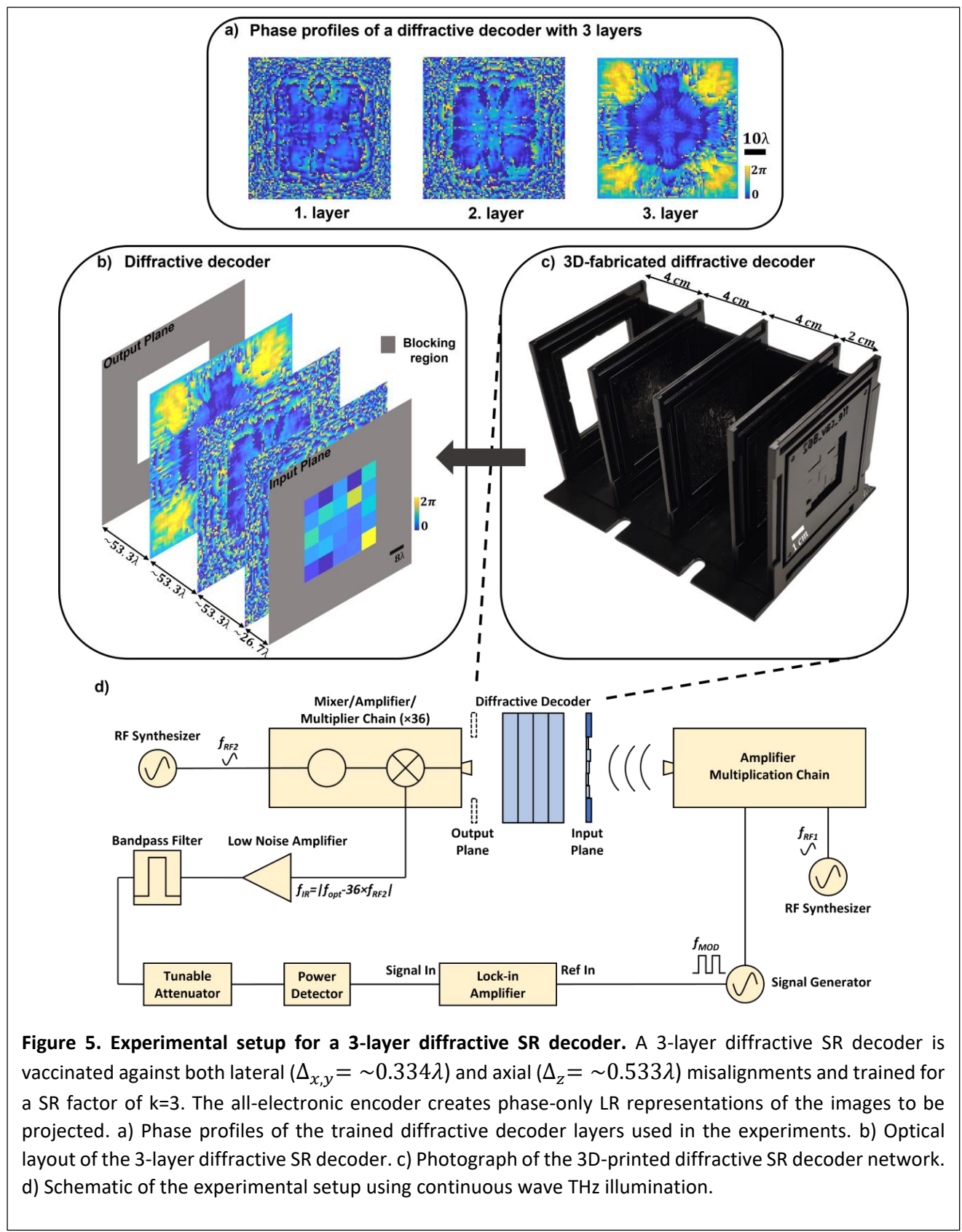

**Figure 5. Experimental setup for a 3-layer diffractive SR decoder.** A 3-layer diffractive SR decoder is vaccinated against both lateral ($\Delta_{x,y}=~0.334\lambda$) and axial ($\Delta_z=~0.533\lambda$) misalignments and trained for a SR factor of k=3. The all-electronic encoder creates phase-only LR representations of the images to be projected. a) Phase profiles of the trained diffractive decoder layers used in the experiments. b) Optical layout of the 3-layer diffractive SR decoder. c) Photograph of the 3D-printed diffractive SR decoder network. d) Schematic of the experimental setup using continuous wave THz illumination.

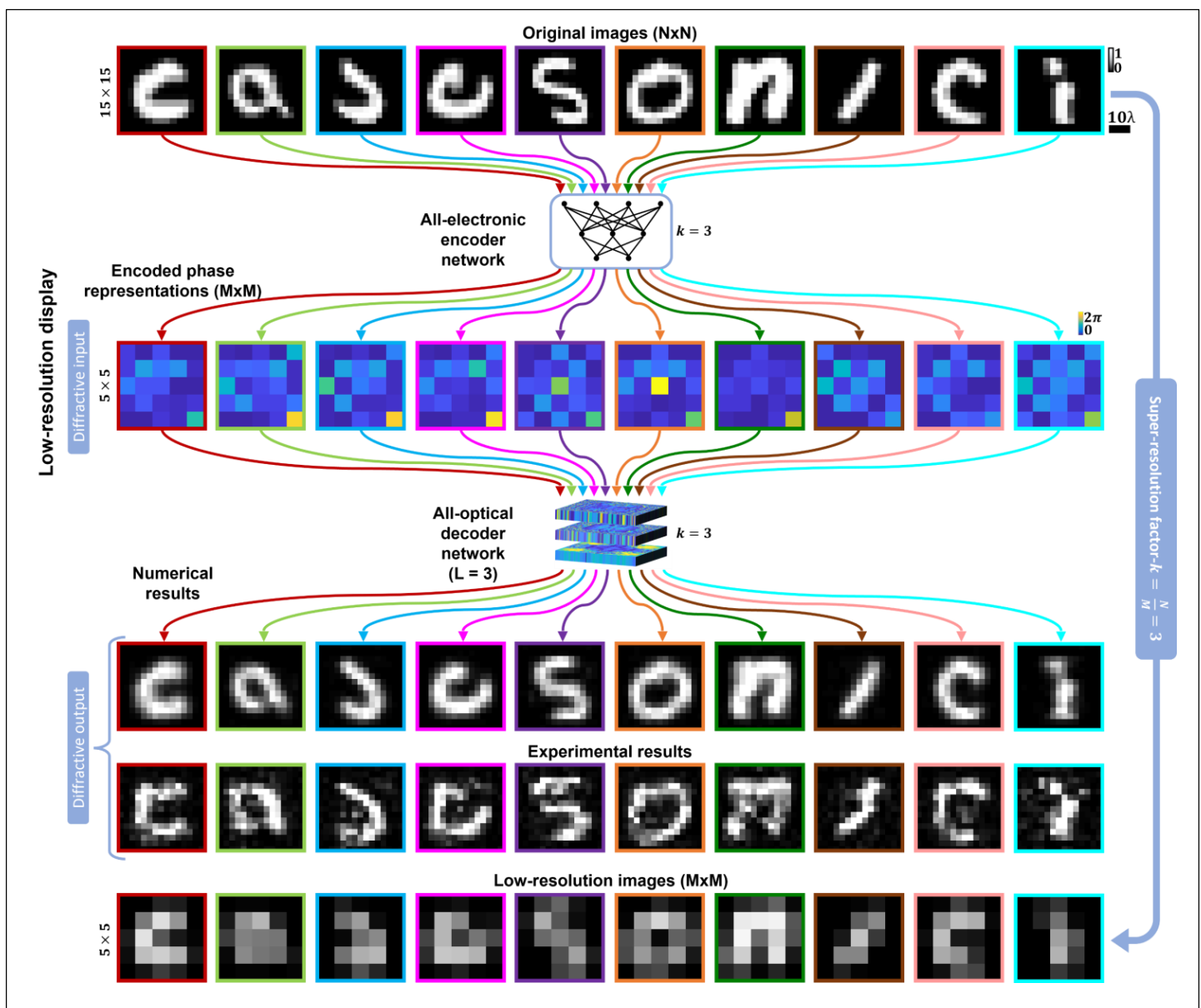

**Figure 6. Experimental results of the diffractive SR image display with L=3 layers.** Encoded phase-only representations of the objects are obtained using the all-electronic encoder. The all-optical diffractive decoder projects super-resolved images. For comparison, low resolution versions of the same images using the same number of pixels as the corresponding wavefront modulator are illustrated at the bottom row of the figure.

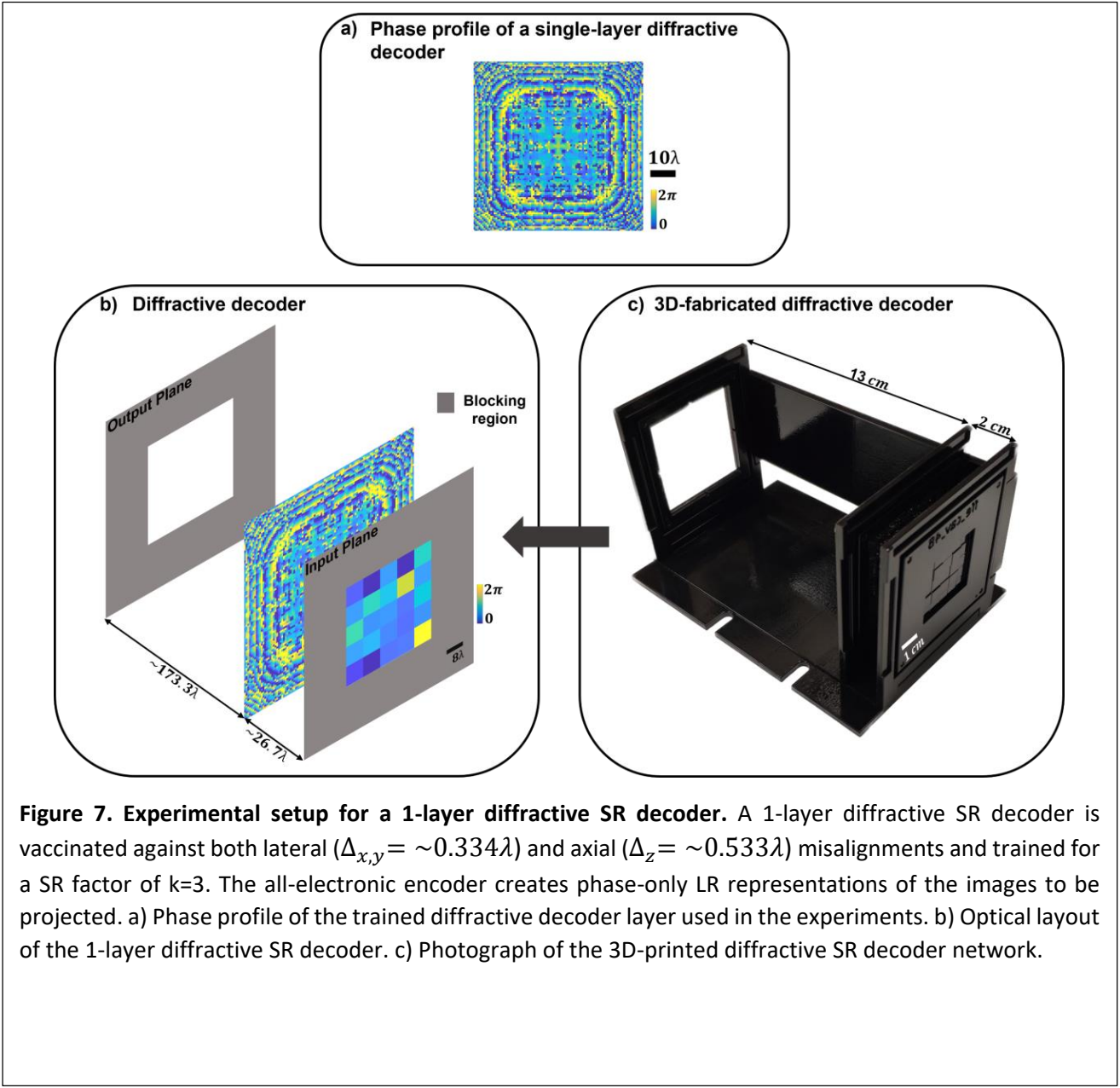

**Figure 7. Experimental setup for a 1-layer diffractive SR decoder.** A 1-layer diffractive SR decoder is vaccinated against both lateral ($\Delta_{x,y} = \sim 0.334\lambda$) and axial ($\Delta_z = \sim 0.533\lambda$) misalignments and trained for a SR factor of k=3. The all-electronic encoder creates phase-only LR representations of the images to be projected. a) Phase profile of the trained diffractive decoder layer used in the experiments. b) Optical layout of the 1-layer diffractive SR decoder. c) Photograph of the 3D-printed diffractive SR decoder network.

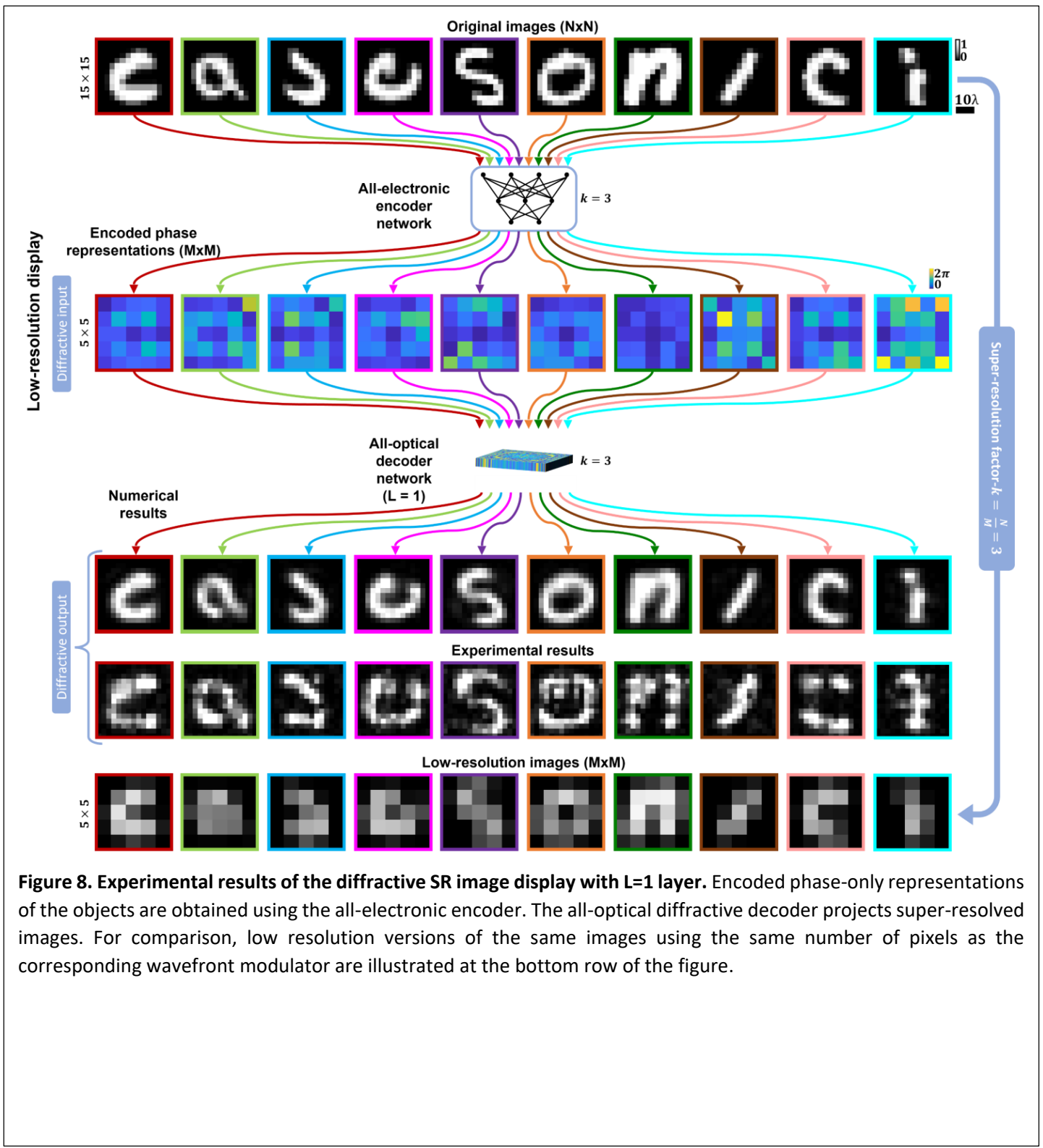

**Figure 8. Experimental results of the diffractive SR image display with L=1 layer.** Encoded phase-only representations of the objects are obtained using the all-electronic encoder. The all-optical diffractive decoder projects super-resolved images. For comparison, low resolution versions of the same images using the same number of pixels as the corresponding wavefront modulator are illustrated at the bottom row of the figure.

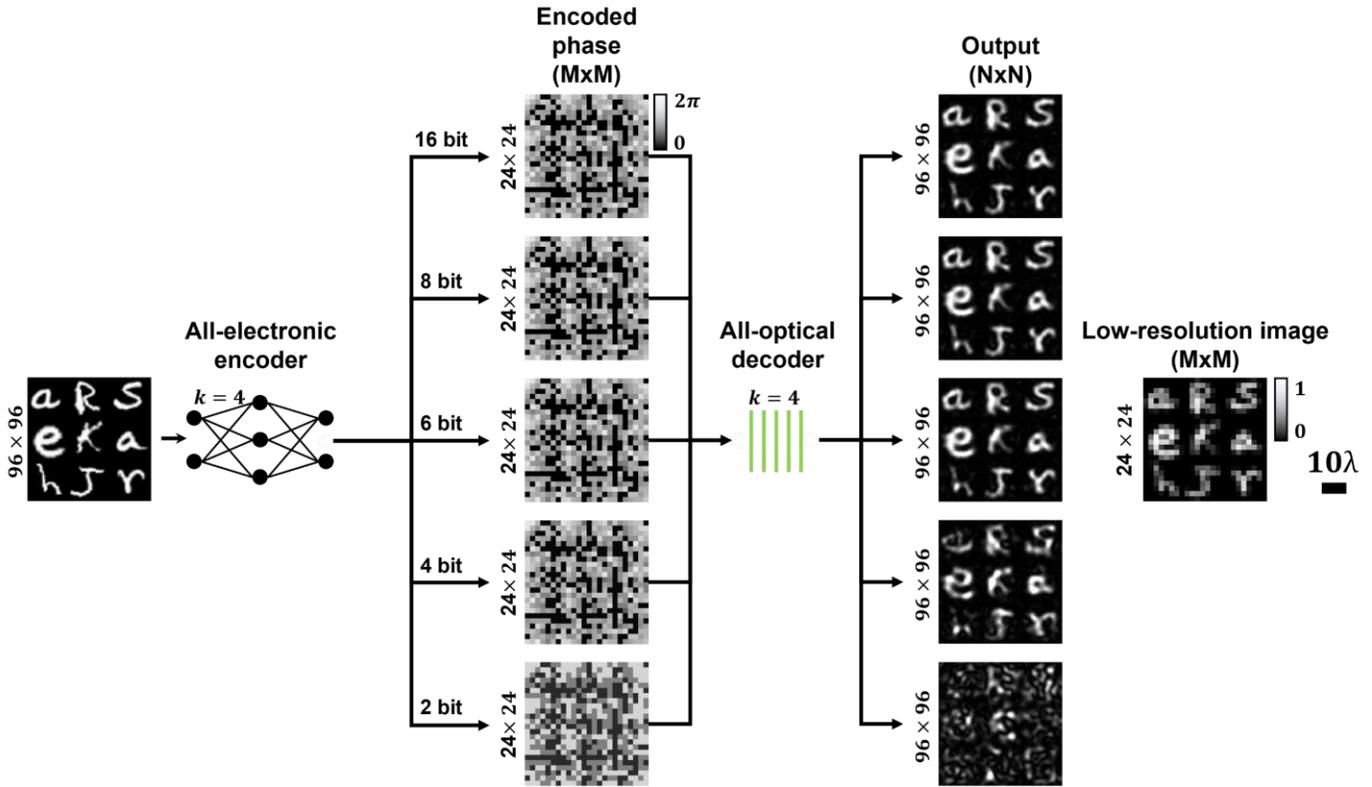

**Figure 9. Quantization analysis of the phase-only wavefront modulation for synthesized images.** Image projection results of the SR display using 5 diffractive layers (L=5) are demonstrated for different phase quantization levels (16-, 8-, 6-, 4-, and 2-bit). The encoding-decoding framework is trained for 16-bit phase quantization of the SLM patterns and blindly tested for lower quantization levels.